\documentclass[10pt]{article} % use larger type; default would be 10pt

\usepackage{cite}
\usepackage{amssymb}
\usepackage{amsmath}
\usepackage{rawfonts}
\usepackage{graphicx}
\usepackage[usenames,dvipsnames]{color}
\usepackage{bbm}

\input prepictex
\input pictex
\input postpictex

\usepackage{cmbright}
\usepackage[T1]{fontenc}
\usepackage{textcomp}

\usepackage{graphicx}
\usepackage{epsfig}
\usepackage{subfigure}

\def\L{\left(}   \def\R{\right)}
   
\def\q{\quad}

\def\LA{\left\langle} \def\RA{\right\rangle}

\def\w#1{\overline{#1}}

\def\C#1{{\cal{#1}}}

\def\Ref#1{(\ref{#1})}

\def\sfrac#1#2{\hbox{\normalsize $\frac{#1}{#2}$}}

\def\gammas#1{{\gamma_#1}}

\def\Lattice{{\mathbb Z}^2}
\def\edge#1#2{{#1}{\sim}{#2}}
\def\vv{\,\hbox{\Large$|$}}

\title{The entropic pressure of a lattice polygon}
\author{F. Gassoumov and E.J. Janse van Rensburg\\
Department of Mathematics and Statistics, York University \\
Toronto, Ontario, M3J 1P3, Canada}
\date{\today} % Activate to display a given date or no date (if empty),
\begin{document}
\maketitle

\begin{abstract}
The entropic pressure in the vicinity of a two dimensional
square lattice polygon is examined as a model of the
entropic pressure near a planar ring polymer.  The scaling
of the pressure as a function of distance from the polygon
and length of the polygon is determined and tested numerically.
\end{abstract}

\section{Introduction}

A polymer confined by a hard wall or other geometrical obstacle
loses conformational entropy.  This loss of entropy causes a net
force on the wall or obstacle \cite{MN91}. These forces have 
been observed experimentally \cite{BDD95,CS95} and modelled 
numerically by using self-avoiding walk models of a 
grafted polymer \cite{DJ13}.

More generally, the physical properties of a polymer in a good solvent are
dependent on its conformational entropy \cite{deG79}.  For 
example, the polymeric stabilisation of a colloid \cite{P91} can be explained
in terms of induced repulsive forces between colloid particles
due to the loss of conformational degrees of freedom of a 
polymer confined between adjacent particles -- see for
example reference \cite{BORW05} for a directed lattice path model.

The forces induced by the loss of conformational entropy in
polymeric systems are the result of an \textit{entropic pressure}
in the vicinity of a polymer.  This pressure has been studied
and simulated in the scientific literature 
-- see for example references \cite{BMJ00,GWF06,MTW77}.

\begin{figure}[h!]
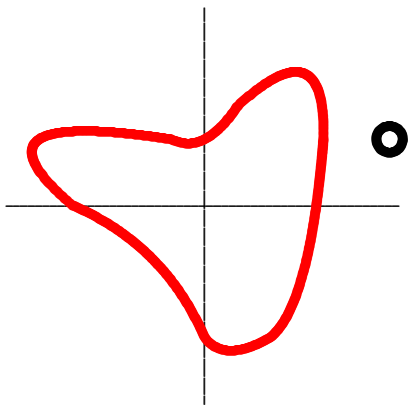

\input figure1.tex
\caption{Schematic illustration of a small test particle near a ring polymer.
The polymer loses entropy so that the particle experiences
an average pressure gradient.  This pressure gradient
causes the particle to be expelled from the vicinity of the
polymer (since it is much lighter than the polymer).}
\label{figure1}    %%ZXZ[figure1]
\end{figure}

A particular simple situation arises when a test particle is placed
close to a polymer.  The polymer will lose conformational degrees 
of freedom -- this implies that the polymer will lose some entropy.
The loss of entropy induces a repulsive force on the test particle, 
which experiences a net pressure gradient in the vicinity of the polymer.  

In figure \ref{figure1} a schematic illustration of a test
particle near a ring polymer is shown.  Close to
the polymer the particle will experience a net pressure, and
a net entropic force along the gradient of the pressure.  

In this paper we examine and characterise the scaling
of the entropic pressure close to a self-avoiding walk
model of a ring polymer.   While our scaling arguments will
be general, our numerical simulations will be of a 
two dimensional lattice model of the situation in figure
\ref{figure1}.

\begin{figure}[t!]
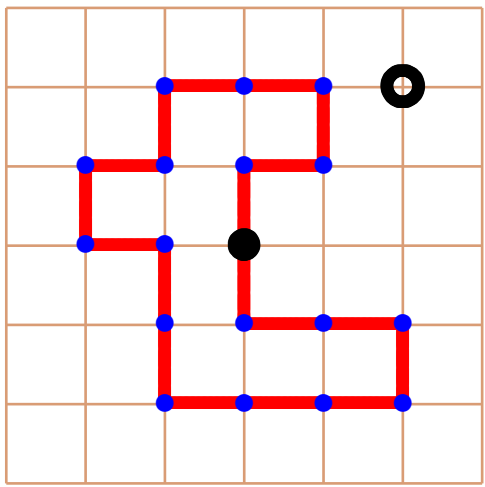

\input figure2.tex
\caption{A lattice polygon model of the ring polymer in figure \ref{figure1}.
The lattice ring polymer is rooted at the origin, and a unit mass
test particle is placed at a lattice site near it.  The presence of the
test particle reduces the conformational degrees of freedom
of the polygon, and thus reducing it entropy.  The loss of entropy induces
a pressure and a net force on the particle along the gradient of the
pressure.}
\label{figure2}    %%ZXZ[figure2]
\end{figure}

Our lattice model is illustrated in figure \ref{figure2}.  A ring polymer is
modelled by a square lattice polygon, which is assumed to be
grafted or fixed at the origin in the lattice.  Since the polymer will
be much heavier than a unit mass test particle placed close to
it, grafting it at the origin amounts to an assumption that the
polygon will not be displaced by an approaching test particle 
(instead, the opposite will happen, the much lighter particle will be
expelled from the vicinity of the polygon by an entropically 
induced repulsive force).

In section 2 the entropic pressure in the vicinity of 
rooted square lattice polygon is examined.  In particular,
a scaling relation for the decay of the pressure with
distance from the polygon is derived.  

Thus, consider a rooted lattice polygon in the square
lattice, and suppose that the number of rooted square lattice
polygons of length $n$ is $p_n$.  Then the conformational entropy of the
polygon will be given by $S_n = k_B \log p_n$, where $k_B$ is 
Boltzmann's constant.  One may check that 
$p_4=4$, $p_6=12$, $p_8=56$, and so on.  

We denote the number of lattice polygons of length $n$ rooted at the
origin and passing through the lattice site $\vec{r}=(x,y)$
by $p_n(\vec{r})$.  Then $ \w{p_n(\vec{r})}
=p_n - p_n(\vec{r})$ is the number of lattice polygons avoiding
the lattice site $\vec{r}$.  The entropy of lattice polygons
avoiding the lattice site $\vec{r}$ is $\w{S_n(\vec{r})}
= k_B \log \w{p_n(\vec{r})}$.

If the lattice site $\vec{r}$ is excluded, then polygons suffer
a loss in entropy given by
\begin{equation}
\Delta\,S_n(\vec{r}) = k_B \log \w{p_n(\vec{r})} - k_B \log p_n,
\label{eqnS}    %%ZXZ[eqnS]
\end{equation}
so that the free energy of the system changes by $\Delta \C{F}(T) = 
- T\,\Delta\,S_n(\vec{r})$.  The \textit{pressure} at $\vec{r}$ can
be defined in terms of the free energy by
\begin{equation}
P_n(\vec{r}) = \frac{\Delta \C{F}(T)}{\Delta V(\vec{r})}
=  -T \, \frac{\Delta\,S_n(\vec{r})}{\Delta V(\vec{r})} .
\end{equation}
Choose units so that $T=k_B=1$ and let $\Delta V(\vec{r})$
be the (unit square or cubical) volume element with $\vec{r}$ at its centre.  By
equation \Ref{eqnS} the pressure at the point $\vec{r}$ is given by
\begin{equation}
P_n(\vec{r}) = \log p_n - \log \w{p_n(\vec{r})}
= - \log \L  1 - \frac{p_n (\vec{r})}{p_n}\R .
\label{eqnP}    %%ZXZ[eqnP]
\end{equation}
By convention the pressure is non-negative, and may be 
considered a particular discrete derivative of the extensive
free energy with respect to a unit change in volume at the
lattice site $\vec{r}$.

\begin{figure}[t!]
\centering
\includegraphics[width=100mm,trim=0 80 30 20]{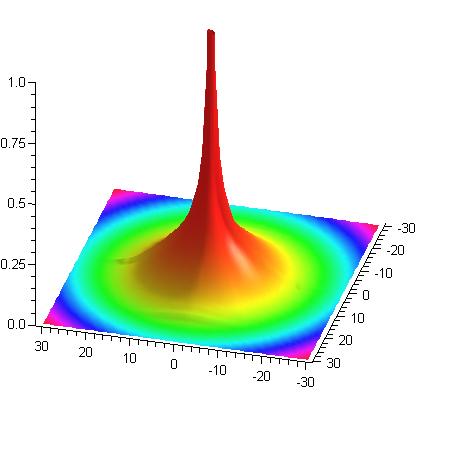}
\caption{The pressure field near a lattice polygon of length $100$
rooted at the origin.  The pressure peaks sharply at the origin,
and decays to zero with distance.  It is also isotropic around the origin.}
\label{figure3d}    %%ZXZ[figure3d]
\end{figure}

For example, if $n=4$, then $p_4 = 4$ and $\w{p_4 (1,0)}=
p_4(1,0) = 2$, so that the pressure at $(1,0)$ is given 
by $P_4(1,0) = - \log \L 1- \sfrac{2}{4} \R = \log 2$.
In figure \ref{figure3d} the pressure field in the vicinity of 
a lattice polygon of length $100$ is illustrated. 

In this paper the scaling of $P_n(\vec{r})$ is examined.  
Let $r = |\vec{r}|$ and let $P_n(r)$ be the mean pressure
a distance $r$ from the origin (averaged over all directions).  

The scaling of $P_n(r)$ will be determined by rescaling
distance in the problem as follows.  The length scale for lattice
polygons is set by the \textit{metric exponent} $\nu$, so that
the mean distance of monomers from the origin on a rooted lattice 
polygon of length $n$ has asymptotic scaling given by
 $\LA r \RA_n \simeq C_0\, n^\nu$ (see for example 
reference \cite{deG79}). 

Define the rescaled pressure $\mathbf{P}_n(a) =
P_n(a\LA r \RA_n)$.  That is, for $a\geq 0$ the pressure
at a distance $a\LA r \RA_n$ from the origin is given by
$\mathbf{P}_n(a)$.

We give a scaling analysis to argue that $\mathbf{P}_n(a)$
scales as follows
\begin{equation}
\mathbf{P}_n(a) \simeq C \, g(a)\,a^{-4}\,n^{-19/32}
\label{Pscaling}    %%ZXZ{Pscaling}
\end{equation}
in two dimensions, for some constant $C$ and where $g(a)$ is a 
function of $a$ which quickly approaches zero as $a$ increases.

In section 3 we present numerical data on the pressure of lattice polygons to
test the above scaling result, by implementing the GAS-algorithm
\cite{JvRR09,JvR10} using BFACF elementary moves \cite{AA83,BF81}.  
Our results verify the $n$-dependence in equation \Ref{Pscaling}.  
Moreover, the pressure is found to be isotropic about the 
origin, decaying at the same rate in any direction with distance
from the origin according to equation \Ref{Pscaling}.  

A unit mass test particle placed near the rooted polygon will
experience a net force due to a pressure gradient.  If the particle
can move freely, then it will accelerate from the vicinity of the
polygon, reaching a terminal velocity if one assumes that it
moves  frictionless without dissipation of energy.  We determine 
the terminal velocity of the particle (assuming that the polygon is 
kept at constant temperature).  Our numerical results show that the terminal
velocity is independent of the size of the polygon, and only 
dependent on its initial position.  For example, it seems
that if a test particle is placed
at the lattice point $(1,0)$, then it will accelerate to the same
terminal velocity for a polygon of any size $n$.

The paper is concluded in section 4 with a few final remarks and 
a summary of our results.

\section{Scaling of the entropic pressure}

Denote the square lattice by $\Lattice$.   A lattice point 
$\vec{v}=(n,m)\in\Lattice$ is a \textit{vertex}.  If two vertices 
$\vec{v}$ and $\vec{w}$ are a unit distance apart, then they are 
the endpoints of an edge $\edge{\vec{v}}{\vec{w}}$.

A \textit{self-avoiding walk} of length $n$ 
from the origin in $\Lattice$ is a sequence of 
edges $\edge{\vec{v}_i}{\vec{v}_{i+1}}$ for $i=0,1,\ldots,n{-}1$,
where the $\vec{v}_i$ are distinct and $\vec{v}_0$ is the origin.  If the walk
is constrained such that $\vec{v}_n = \vec{v}_0$, then the walk is a 
\textit{lattice polygon} of length $n$ (and it is rooted at the origin).
Notice that the lattice polygon can be oriented in two ways from the
origin, but that by convention it is not oriented.

The number of self-avoiding walks of length $n$ (steps or edges)
from the origin is denoted by $c_n$.   It is known that 
$\lim_{n\to\infty} \sfrac{1}{n} \log c_n = \log \mu$ \cite{HM54}
and $\log \mu$ is the \textit{connective constant} while
$\mu$ is the \textit{growth constant} of the self-avoiding walk.

The number of self-avoiding walks is believed to have asymptotic
behaviour
\begin{equation}
c_n \simeq A\,n^{\gamma-1}\,\mu^n
\label{eqnA1}   %%ZXZ[eqnA1]
\end{equation}
where $\gamma$ is the entropic exponent of the self-avoiding walk
and has exact value $\gamma = \sfrac{43}{32}$
in two dimensions  \cite{D86}.

Below we shall be concerned with walks from the origin in a 
\textit{half-space}.  Let $L$ be a line through the origin in 
${\mathbb R}^2$.  Then $L$ cuts $\Lattice$ into two 
\textit{half-lattices}.  This generalises in higher dimensions where
$L$ is a hyperplane of dimension $d{-}1$.

The number of self-avoiding walks of length $n$ from the origin 
confined to one of the half-lattices will be denoted by $c_n^+(L)$.
The asymptotic form for $c_n^+(L)$ is believed to be given by
\begin{equation}
c_n^+(L) \simeq A^\prime\, n^{\gamma_1-1}\,\mu^n,
\label{eqnA2}   %%ZXZ[eqnA2]
\end{equation}
where $\gamma_1= \sfrac{61}{64}$ in two dimensions (see for example
reference \cite{C87}).

Denote the number of lattice polygons of length $n$ rooted at the 
origin by $p_n$.  Then it is known that $\lim_{n\to\infty}
\sfrac{1}{n} \log p_n = \log \mu$ \cite{H61,HW62}, where
the limit is taken through even integers.

It is expected that $p_n$ has asymptotic behaviour given by
\begin{equation}
p_n \simeq B\, n^{\alpha_s-2}\,\mu^n
\label{eqnB3}   %%ZXZ[eqnB3]
\end{equation}
where $\alpha_s$ is the entropic exponent of lattice polygons
and has exact value $\alpha_s = \sfrac{1}{2}$ in two dimensions; 
see for example reference \cite{D90}.

\subsection{The length scale}

Let $c_n(\vec{r})$ be the number of self-avoiding walks 
in $\Lattice$ of length $n$ steps from the origin to the lattice point 
$\vec{r}$. 

Similarly, let $c_n(r) = \sum_{r=|\vec{r}|} c_n(\vec{r})$ be the 
number of self-avoiding walks of length $n$ from the origin
to lattice points a distance $|\vec{r}|=r$ on a spherical shell
from the origin. Since the endpoints of walks counted by $c_n(r)$ ends 
on a spherical shell a distance $r$ from the origin, one should have
\begin{equation}
c_n(r)  \simeq A_0\, r^{d-1}\, c_n(\vec{r}) 
\label{eqnA3}   %%ZXZ[eqnA3]
\end{equation}
in $d$ dimensions.

One may similarly consider walks in a half-space instead.  Let
$c_n^+(\vec{r};L)$ be the number of walks from the origin
in a half-space defined by the $(d{-}1)$-dimensional
hyperplane $L$, of length $n$ and ending
in the vertex $\vec{r}$.  Put $c_n^+(r;L) 
= \sum_{r=|\vec{r}|} c_n^+(\vec{r};L)$, the number of 
walks from the origin in a half-space defined by $L$, of length 
$n$ and with endpoint a distance $r$ from the origin.  Since
these walks end on a hemispherical shell of radius $r$ from the
origin, one should have
\begin{equation}
c_n^+(r;L)  \simeq A_1\, r^{d-1}\, c_n^+(\vec{r};L) 
\label{eqnA33}   %%ZXZ[eqnA33]
\end{equation}
in $d$ dimensions.

The average distance of the endpoint of a self-avoiding walk of
length $n$ from the origin is denoted by $\LA r \RA_n$, and this
introduces a metric in the model of self-avoiding walks.
It is expected that $\LA r \RA_n \simeq C_0\, n^{\nu}$, where
$\nu$ is the \textit{metric exponent}.  In two dimensions the
exact value is $\nu = \sfrac{3}{4}$ \cite{N82,N84}.

In terms of $c_n(r)$ the average distance of the endpoint of the
walk from the origin may be calculated from 
\begin{equation}
\LA r \RA_n = \frac{\sum_{r\geq 0} r\, c_n(r)}{c_n} \simeq C_0\, 
n^\nu .
\label{eqnrnu}   %%ZXZ[eqnrnu]
\end{equation}
This gives a way for estimating the scaling of $c_n(r)$.

For fixed $n$, $c_n(r)$ should decay as $r$ increases with length 
scale $n^\nu$.  That is, one may guess that 
\begin{equation}
c_n(r) = B_0\, r^x\,e^{-r/C_0 n^\nu} c_n
\label{eqncnrc}   %%ZXZ[eqncnrc]
\end{equation}
where $x$ is an exponent that must be determined.  This 
assumption may be used to determine the scaling of $\LA r \RA_n$.
Substituting this assumption in equation \Ref{eqnrnu} and
approximating the summation by an integral and using 
equation \Ref{eqncnrc} gives 
\begin{equation}
\LA r\RA_n  % \simeq \frac{\sum_{r\geq 0} r\,c_n(r)}{c_n}
\simeq \int_0^n B_0\, r^{1+x} e^{-r/C_0 n^\nu} dr
\simeq B_0 \left( C_0 n^\nu \right)^{x+2}  \int_0^\infty
\left[ \frac{r}{C_0n^\nu} \right]^{x+1} e^{-r/C_0 n^\nu} 
\frac{dr}{C_0 n^\nu} \nonumber
\end{equation}
assuming that $n\gg n^\nu$ (since $\nu=\sfrac{3}{4}$ in two dimensions,
this is true for large $n$).  The last integral is a constant, so 
one concludes that
\begin{equation}
\LA r\RA_n \sim  n^{\nu(x+2)} .
\end{equation}
Since $\LA r \RA_n \simeq C_0\, n^{\nu}$,
it follows that $x=-1$ in equation \Ref{eqncnrc}.

Taking the above together shows that $c_n(r)$ scales as
\begin{equation}
c_n(r) \simeq B_0\, r^{-1}\, e^{-r/C_0 n^\nu}\, c_n .
\label{eqncnrcA}   %%ZXZ[eqncnrcA]
\end{equation}
This result gives the scaling of $c_n(\vec{r})$ as well by equation
\Ref{eqnA3}:
\begin{equation}
c_n(\vec{r}) \simeq \frac{r^{1-d}}{A_0}\, c_n(r) \simeq
\frac{B_0}{A_0 r^d}\, e^{-r/C_0 n^\nu} c_n  
= A\, r^{-d}e^{-r/C_0 n^\nu} c_n 
\label{eqncnrcBA}   %%ZXZ[eqncnrcBA]
\end{equation}
where $r = |\vec{r}|$ and $A$ is a constant.

The above arguments apply mutatis mutandis to $c_n^+(\vec{r})$,
with $c_n$ in the above replaced by $c_n^+$.  This follows in
particular because the scaling of the average distance of the endpoint
of half-space walks is given by
\begin{equation}
\LA r \RA_n^+ 
= \frac{\sum_{r\geq 0}r\,c_n^+(r)}{c_n^+} 
\simeq C_1 \, n^\nu .
\label{eqnrnuplus}   %%ZXZ[eqnrnuplus]
\end{equation}
The rest of the argument is the same.  That is, one finally obtains
\begin{equation}
c_n^+(\vec{r},L) \simeq \frac{r^{1-d}}{A_1}\, c_n^+(r) \simeq
\frac{B_1}{A_1 r^d}\, e^{-r/C_1 n^\nu} c_n^+ 
= B\, r^{-d}e^{-r/C_1 n^\nu} c_n^+ (L)
\label{eqncnrcB}   %%ZXZ[eqncnrcB]
\end{equation}
where $r = |\vec{r}|$ and $B$ is a constant.

\subsection{The number of polygons passing through a lattice point $\vec{r}$}

To determine a scaling form for the pressure due to a rooted 
lattice polygon, it will be necessary to determine a scaling relation
for lattice polygons passing through a lattice point $\vec{r}$.

Denote the number of rooted lattice polygons of length $n$
passing through the point $\vec{r}$ by $p_n(\vec{r})$.  
Approximate this by considering a pair of walks which step from 
the origin to $\vec{r}$.  If the walks avoid one another, then 
their union is a rooted lattice polygon passing through $\vec{r}$.  
If they do not avoid one another, then an upper bound on $p_n(\vec{r})$ 
is obtained.  

However, if the two walks are confined to two different half-spaces 
defined by a $(d{-}1)$-dimensional
hyperplane $L$ passing through the origin and $\vec{r}$, then
they avoid one another, and one may estimate 
\begin{equation}
 p_n(\vec{r}) \gtrsim \sum_{k=0}^n c_k^+(\vec{r},L)\, c_{n-k}^+(\vec{r},L) .
\label{eqn14B}   %%ZXZ[eqn14B]
\end{equation}
This may be approximated by using equation \Ref{eqncnrcB} as
estimates for $c_n^+(\vec{r},L)$.  This gives
\begin{equation}
 p_n(\vec{r}) \gtrsim B^2 \sum_{k=0}^n r^{-2d}
e^{-r/C_1 k^\nu} e^{-r/C_1 (n-k)^\nu} 
c_k^+(L)\,c_{n-k}^+(L) .
\label{eqn14C}   %%ZXZ[eqn14C]
\end{equation}
Next, rescale $r = |\vec{r}|$ by its mean $\LA r \RA_n^+
\simeq C_1\, n^\nu$:  That is, choose $r=a\,\LA r \RA_n^+$
in the above, for $a>0$ in equation \Ref{eqn14C}.  This
estimates the number of polygons through the point
$\vec{r} = \L a\,\LA r \RA_n^+\R\, \sfrac{\vec{r}}{|\vec{r}|}$.  Denote this
by $\widehat{p}_n(a,L)$ (that is, $\widehat{p}_n(a,L)$ is the
number of polygons of length $n$ passing through a point
fixed on $L$ a distance $\rho = a\LA r \RA_n^+$ from the origin).
This gives
\begin{equation}
 \widehat{p}_n(a,L) \gtrsim B^2
\sum_{k=0}^n 
\L a\, C_1\, n^\nu \R^{-2d}
e^{-a(n/k)^\nu} e^{-a(n/(n-k))^\nu}
c_k^+(L)\, c_{n-k}^+ (L) .
\end{equation}
The scaling of $c_n^+(L)$ is given in equation \Ref{eqnA2},
and is asymptotically independent of $L$.  Thus, the above may
be averaged over $L$ (that is, over all directions through the 
origin) so that the number of polygons passing through a point
a distance $\rho = a\LA r \RA_n^+$ from the origin is approximated
by
\begin{align}
\widehat{p}_n(a) &\gtrsim 
\frac{B^2(A^\prime)^2\mu^n}{\L a C_1 n^\nu \R^{2d}}
\int_0^n k^{\gammas1-1}\,(n-k)^{\gammas1-1}\,
e^{-a(n/k)^\nu} e^{-a(n/(n-k))^\nu}\, dk \nonumber \\
&= \frac{B^2(A^\prime)^2\mu^n}{\L a C_1 n^\nu \R^{2d}}
\, n^{2\gammas1-1}\,\int_0^1 x^{\gammas1-1}(1-x)^{\gammas1-1}
e^{-a/x^\nu}\,e^{-a/(1-x)^\nu}\, dx 
\end{align}
where the summation is now approximated by an integral.

Denote the integral above by $g(a)$:
\begin{equation}
g(a) = \int_0^1  x^{\gammas1-1}(1-x)^{\gammas1-1}
e^{-a/x^\nu}\,e^{-a/(1-x)^\nu}\, dx .
\label{eqn17BC}   %%ZXZ[eqn17BC]
\end{equation}
Put $C_2 = B^2(A^\prime)^2/C_1^{2d}$.  Then the above reduces to
\begin{equation}
\widehat{p}_n(a) \gtrsim C_2\, \mu^n\, a^{-2d}g(a)\, n^{2\gammas1-1-2d\nu}.
\label{eqn17B}   %%ZXZ[eqn17B]
\end{equation}
Notice that the dependence on direction has disappeared, so that
$\widehat{p}_n(a)$ is the approximate number of rooted polygons of length $n$, 
passing through a particular point a distance equal to $\rho = 
a \LA r\RA_n^+ \simeq a\, C_1\,n^\nu$ from the origin.

\begin{figure}[t!]
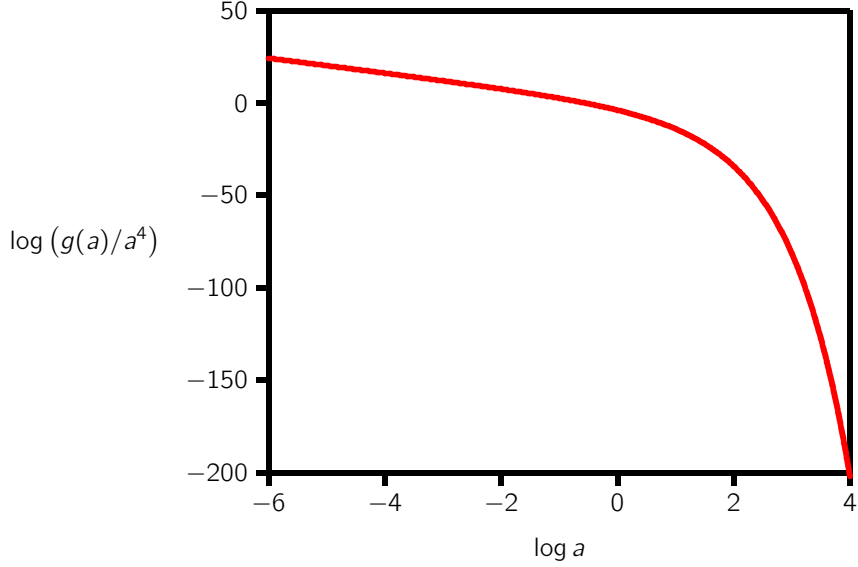

\input figureG.tex
\caption{A plot of $\log\L g(a)\,a^{-4} \R$ against
$\log a$.}
\label{figureG}    %%ZXZ[figureG]
\end{figure}

If $d=2$, then $\gammas1=\sfrac{61}{64}$ and $\nu=\sfrac{3}{4}$, 
and the $a$- and $n$-dependence of $\widehat{p}_n(a)$ becomes
\begin{equation}
\widehat{p}_n(a) \gtrsim \frac{C_2\,\mu^n g(a)}{a^4} \, n^{-67/32}.
\label{eqn24}   %%ZXZ[eqn24]
\end{equation}
It remains to determine the pressure.  A plot of $\sfrac{g(a)}{a^4}$ is given
in figure \ref{figureG} on a log-log scale.

\subsection{The rescaled pressure $\mathbf{P}_n(a)$}

By equation \Ref{eqnP} the pressure due to rooted polygons at a point
a distance $\rho = a\,\LA r \RA_n^+ \simeq  a\, C_1\,n^\nu $ from the origin is 
approximately given by
\begin{equation}
\mathbf{P}_n(a) = - \log \L 1- \frac{\widehat{p}_n(a)}{p_n}\R,
\end{equation}
where $\widehat{p}_n(a)$ is the number of polygons passing 
through a point a distance $\rho = a\LA r \RA_n^+$ from the origin.

The scaling of this is obtained by using equations \Ref{eqnB3} and 
\Ref{eqn17B}.  In particular, for $a$ and $n$ large this becomes
\begin{equation}
\mathbf{P}_n(a)  \simeq - \log \L 1 - \frac{C_2\, g(a)}{B\, a^{2d}}
\, n^{2\gammas1-1-2d\nu-\alpha_s + 2} \R .
\end{equation}
It is readily checked that the power of $n$ in the above is 
negative and that $g(a)$ quickly approaches zero with increasing
$a$.  Thus, for large $n$ the logarithm can be expanded.
Keeping only the leading term shows that
\begin{equation}
\mathbf{P}_n(a) \simeq \frac{C_2\, g(a)}{B\, a^{2d}}
\, n^{2\gammas1-1-2d\nu-\alpha_s + 2} .
\label{eqnPscaleG}    %%ZXZ[eqnPscaleG]
\end{equation}
If $d=2$, then substitution of the exact values of the exponents
and combining the constants above give
\begin{equation}
\mathbf{P}_n(a) \simeq \frac{C\,g(a)}{a^{4}}\, n^{-19/32} ,
\label{eqnPscale}    %%ZXZ[eqnPscale]
\end{equation}
for some constant $C$.

This prediction may be tested numerically by plotting
$[n^{19/32}\, \mathbf{P}_n(a)]$ against $[|\vec{r}|/n^{3/4}]$.
That is, rescaling length in the lattice by a factor $n^{\nu}$
and the pressure by $n^{19/32}$ should collapse data for a
range of choices of $\vec{r}$ and $n$ to a single universal 
curve which is only a function of the parameter $a$.  The shape of
this universal curve is given by a rescaling of $\sfrac{g(a)}{a^4}$.

\subsection{The pressure gradient and velocity of a test particle}

A unit mass test particle placed near the origin in the lattice will experience
a average pressure gradient in the vicinity of the polygon.  If the test particle
can move freely without friction, then it will accelerate from high 
to low pressure.

Thus, assume that a particle is placed near the polygon, and that
it will be accelerating in a frictionless environment down the pressure 
gradient.  Assume in addition that temperature is constant (that is, 
the polygon is in contact with a large heat bath keeping its 
temperature fixed).  This last assumption is important since 
the particle will drain energy from the polygon, cooling
it down, in the absence of a heat bath.

Suppose that the particle will move along the $X$-axis.
The pressure gradient between lattice sites $(x,0)$ and $(x+1,0)$
is $\Delta P_n(x,0) = P_n(x+1,0) - P_n(x,0)$.  Assuming that
the acceleration of the particle from $(x,0)$ to $(x+1,0)$ 
is constant, it will be given by
\begin{equation}
\frac{\Delta v_n(x,0)}{\Delta t} = P_n(x,0) - P_n(x+1,0) ,
\label{eqn29}    %%ZXZ[eqn29]
\end{equation}
where $\Delta v_n(x,0) = v_n(x+1,0)-v_n(x,0)$ and $v_n(x,0)$ 
is the speed of the particle at the point $(x,0)$, and $\Delta t$ 
is the time interval it takes for the particle to move from $(x,0)$ to $(x+1,0)$.

Notice that $\sfrac{\Delta v_n(x,0)}{\Delta t}
= \sfrac{\Delta v_n(x,0)}{\Delta x} \sfrac{\Delta x}{\Delta t}
\approx \sfrac{v_n(x,0)\, \Delta v_n(x,0)}{\Delta x}$ since
$ \sfrac{\Delta x}{\Delta t}\approx v_n(x,0)$.  That is, if
$\Delta P_n(x,0) = P_n(x+1,0)-P_n(x,0)$, then equation \Ref{eqn29}
becomes
\begin{equation}
v_n(x,0)\,\Delta v_n(x,0) \approx - \Delta P_n(x,0)
\label{eqn30}     %%ZXZ[eqn30]
\end{equation}
and in particular where $\Delta x = 1$ was used and where
$\Delta P_n(x,0) =   P_n(x+1,0) - P_n(x,0)$ is the pressure
drop from $(x,0)$ to $(x+1,0)$.

Equation \Ref{eqn30} may be approximated by a differential
equation in $x$.  Integrating this from $y$ to $x$ gives
\begin{equation}
\sfrac{1}{2}\,v_n^2(z,0) \vv_y^x = - P_n(z,0)\vv_y^x .
\end{equation}
This becomes
\begin{equation}
\sfrac{1}{2}\, \L v_n^2(x,0)-v_n^2(y,0) \R = P_n(y,0)-P_n(x,0) . 
\label{eqn31}     %%ZXZ[eqn31]
\end{equation}

Observe that
the left hand side is the difference in kinetic energy of a unit mass
particle, and the right hand side is the drop in pressure.
For example, putting $y=1$ gives 
\begin{equation}
v_n(x,0) = \sqrt{v_n^2(1,0) + 2\,\L P_n(1,0) - P_n(x,0)   \R} .
\end{equation}
That is, if the particle is put with zero velocity at $(1,0)$, then
its velocity at $(x,0)$ is
\begin{equation}
v_n(x,0) = \sqrt{2\,\L P_n(1,0) - P_n(x,0)   \R} .
\label{eqn34}     %%ZXZ[eqn34]
\end{equation}
If $x$ is large (say $x>n$) then $P_n(x,0)=0$.  This
gives the \textit{terminal velocity} of the particle once it has moved
far from the polygon:
\begin{equation}
v_n^{ter} = \sqrt{2\, P_n (1,0)}
\label{eqn35}     %%ZXZ[eqn35]
\end{equation}
for a particle released at $(1,0)$ in rest.  

{\renewcommand{\baselinestretch}{1.5}
% Estimates of p_n
\begin{table}[b!]
\begin{center}
\caption{Numerical estimates of $p_n$ and $\w{p_n(1,0)}$}
\label{Tabledata}   %%ZXZ[Tabledata]

\begin{tabular}{||r|ll|rr||}
\hline
$n$\q &\q $p_n$ &\q $\sigma_n$ &\q $\w{p_n(1,0)}$ &\q $\sigma_n$ \\
\hline\hline
$4$      &$4$   &$0$ &$2$ &$0$ \\
$6$      &$12.0002$   &$0.0013$ &$5.0007$ &$0.0014$ \\
$8$      &$56.003$   &$0.014$ &$22.0033$ &$0.0077$ \\
$10$    &$279.978$   &$0.090$ &$109.017$ &$0.046$ \\
$12$    &$1487.56$   &$0.59$ &$582.03$ &$0.29$ \\
$14$    &$8229.7$   &$3.6$ &$3238.51$ &$1.8$ \\
$16$    &$46998$   &$24$ &$18601$ &$11$ \\
$18$    &$274830$   &$160$ &$109315$ &$74$ \\
$20$    &$1636900$   &$1100$ &$654170$ &$490$ \\
\hline
\end{tabular}

\end{center}
\end{table}
}

Notice that the rescaling of $\mathbf{P}_n(a)$ in equation
\Ref{eqnPscaleG} is rotationally symmetric about the origin
in the square lattice (independent of direction).  Hence, the
arguments above generalises to particles accelerating in other
directions from the origin -- it is therefore 
appropriate to consider a particle moving freely along the $X$-axis, 
without loss of generality.  For example, the terminal velocity
will only be dependent on the distance the particle is released
from the origin, and not on the direction of its trajectory.

\begin{figure}[t!]
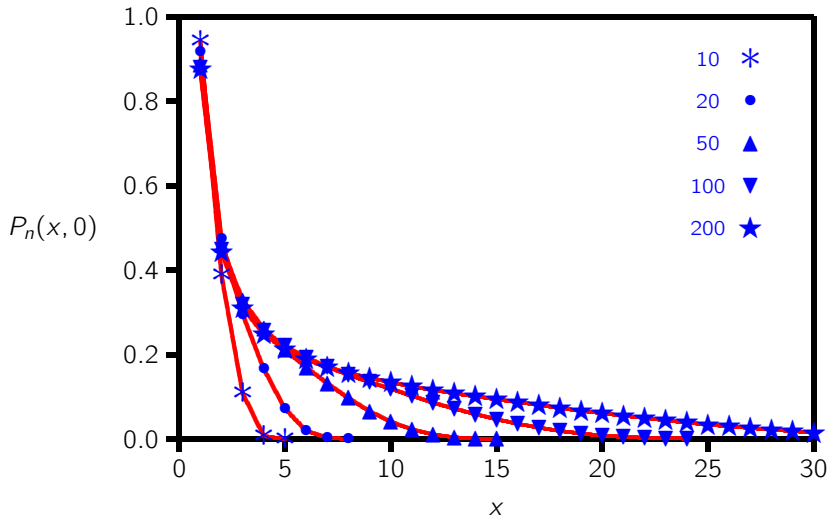

\input figureA.tex
\caption{The pressure $P_n(x,0)$ (equation \Ref{eqnP})
along the $X$-axis at the points $(x,0)$  for $x=1,2,3,\ldots,30$,
and for $n=10$(\textasteriskcentered), $n=20$($\bullet$),
$n=50$($\blacktriangle$), $n=100$($\blacktriangledown$) and
$n=200$($\bigstar$).
}
\label{figureA}    %%ZXZ[figureA]
\end{figure}

\section{Numerical Results}

Rooted lattice polygons can be approximately enumerated \cite{JvR10} 
by using the GAS-algorithm \cite{JvRR09}.  This algorithm was
analysed for polygons in cubic lattices in reference \cite{JvRR11},
and approximate enumeration of lattice polygons were done
in references \cite{JvRR11A,JvRR11B} using BFACF-style 
elementary moves \cite{BF81,AA83}. See for example
reference \cite{MS93} for the application of the BFACF-algorithm
to square lattice self-avoiding walks and polygons.

In this paper the GAS algorithm is used to approximately enumerate
rooted lattice polygons avoiding fixed vertices $\vec{r}$
in the square lattice.  GAS-sampling was 
tuned to be uniform over the lengths $n$ of polygons for $4\leq n
\leq 250$.  Sampling was done along $500$ independent sequences, each
of length $3\times 10^7$.  This gives a total of $1.5 \times 10^{10}$
iterations.  

Results from the simulations are the estimates of the number of 
rooted lattice polygons $p_n$, and the number of rooted lattice
polygons avoiding the lattice site $(x,y)$, denoted by 
$\w{p_n (x,y)}$.  A sample of our results are listed in table
\ref{Tabledata} -- the number of rooted polygons of length $n$, 
and the number of rooted polygons avoiding the lattice site $(1,0)$
are listed, together with their confidence intervals.  

\begin{figure}[t!]
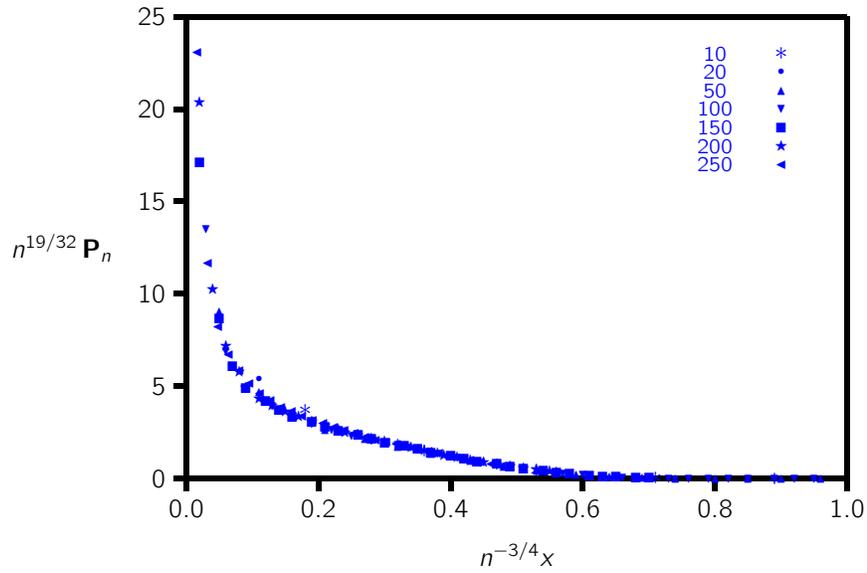

\input figureB.tex
\caption{Testing the scaling prediction in equation 
\Ref{eqnPscale}.  The rescaled pressure $n^{19/32}P_n(x,0)$ 
plotted as a function of $n^{-4/3} x$.  These data include 
all the data points in figure \ref{figureA}.  The data collapse to
a single curve, uncovering the scaling function $g(a)/a^4$
in equation \Ref{eqnPscale}.}
\label{figureB}    %%ZXZ[figureB]
\end{figure}
\begin{figure}[th!]
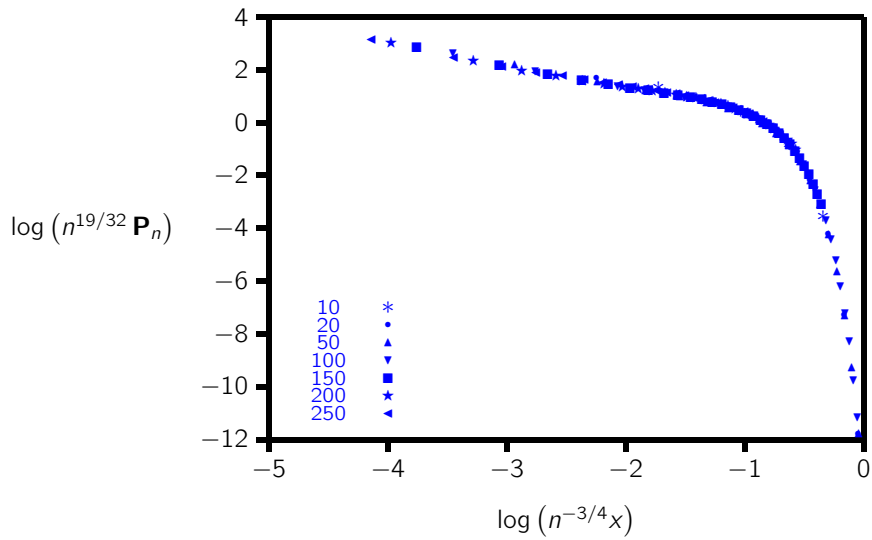

\input figureC.tex
\caption{The same data as in figure \ref{figureB}, but on a 
log-log scale.  These data accumulate along a curve which is
similar in shape to $\frac{g(a)}{a^4}$ plotted in figure
\ref{figureG}, but with rescaled axes.
}
\label{figureC}    %%ZXZ[figureC]
\end{figure}

\begin{figure}[t!]
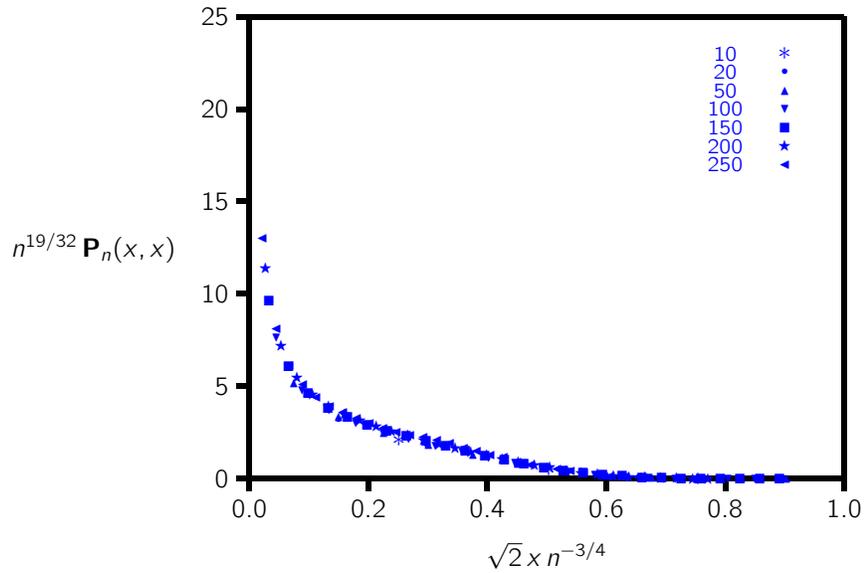

\input figureD.tex
\caption{Testing the scaling prediction in equation 
\Ref{eqnPscale}.  The rescaled pressure $n^{19/32}P_n(x,x)$ 
plotted as a function of $\sqrt{2}\,x\,n^{-3/4}$.   The data collapse to
a single curve, uncovering the scaling function $a^{-4}g(a)$
in equation \Ref{eqnPscale}.}
\label{figureD}    %%ZXZ[figureD]
\end{figure}
\begin{figure}[h!]
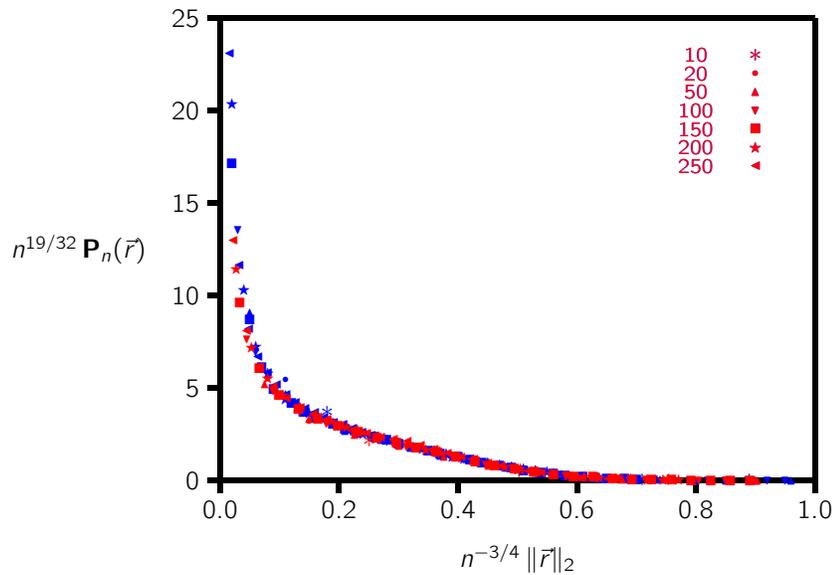

\input figureE.tex
\caption{The data in figures \ref{figureB} and \ref{figureD} plot
on the same scale and axes.}
\label{figureE}    %%ZXZ[figureE]
\end{figure}

\subsection{The pressure}

The pressure as a function of $(x,0)$ for $1\leq x \leq 30$
is plotted for values of $n$ ranging from $n=10$ to $n=200$
in figure \ref{figureA}.  As expected, the pressure 
approaches zero with increasing distance from the 
origin -- for polygons of length $n$ the pressure at lattice
points $(x,y)$ with $|x|>\sfrac{n}{2}$ or $|y|>\sfrac{n}{2}$
is necessarily zero, since a rooted polygon cannot pass through
such points.

The scaling of the pressure can be uncovered by considering
equation \Ref{eqnPscale}.  That is, if distance is rescaled by
$n^{-3/4}$, then pressure should be rescaled by $n^{19/32}$.
This implies that plotting $n^{19/32} P_n(x,0)$ against
$x/n^{3/4}$ should collapse the data in figure \ref{figureA}
onto a single curve.  This is illustrated in figure \ref{figureB}.

The data in figure \ref{figureB} includes points values of 
$n\in\{10,20,50,100,150,200,250\}$ for 
$1\leq x \leq 29$, and show that the scaling of $P_n(x,0)$
is well described by equation \Ref{eqnPscale}.  In fact, the 
approximation made in equation \Ref{eqn14B}, and carried
through equation \Ref{eqn24} to equation \Ref{eqnPscale}
apparently captures the dominant contributions to the pressure.

Plotting the data in figure \ref{figureB} on a log-log scale
gives figure \ref{figureC}.  This plot shows that there are two
scaling regimes.  The first is the data for large $n$ (and $x$
small compared to $n$).  The rescaled pressure is large and
decays slowly with increasing $n$.  The other regime is for large
$x$ (and $n$ small compared to $x$).   In this regime the
pressure decays quickly with distance.  Observe that the shape
of this data is a scaled representation of the plot of $g(a)/a^4$
on a log-log scale in figure \ref{figureG}.  This is not accidental,
as seen in equation \Ref{eqnPscale}. 

\begin{figure}[t!]
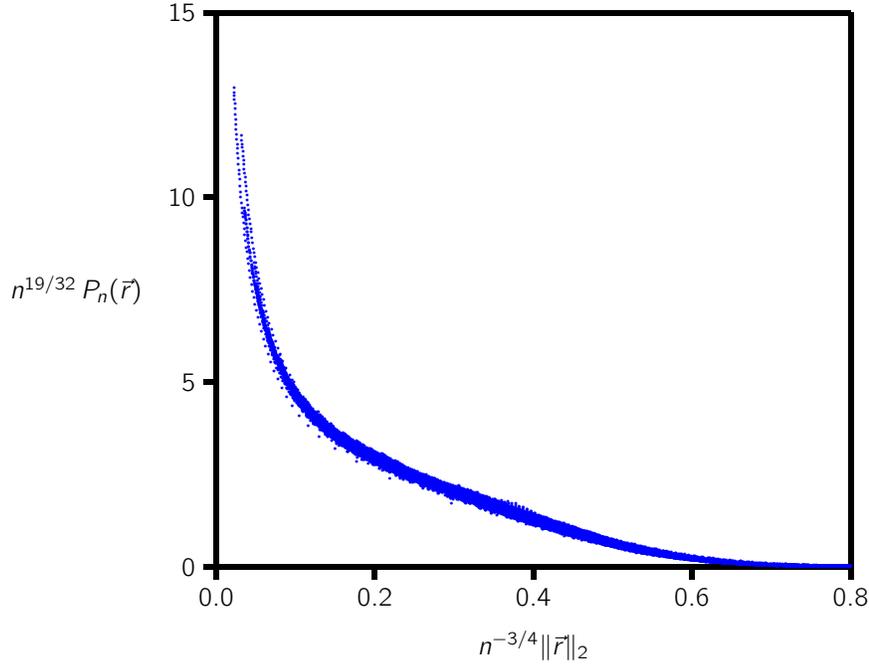

\input figureF.tex
\caption{All the data except those along the $X$-axis.  Each data
point is the rescaled pressure $n^{19/32}P_n(x,y)$ plotted against
$n^{-3/4}\| (x,y) \|_2$ for $x\in [1,29]$, $y\in [1,29]$ and
$n\in [4,250]$ even.  The data accumulate along a single
curve, uncovering the scaling of the pressure field in the vicinity
of a ring polymer.}
\label{figureF}    %%ZXZ[figureF]
\end{figure}

\subsection{Scaling in other directions}

The scaling of ${\mathbf P}_n(a)$ in equation \Ref{eqnPscale} predicts
that the scaling of the pressure is isotropic -- that is, the same
in all direction.  Above, the scaling along a lattice axis
was examined, in this section other directions will be considered.

Consider first the pressure $P_n(x,x)$ at lattice points 
$\{(x,x)\vv 1\leq x \leq 29\}$ along the diagonal direction in the lattice.  The pressure is 
a function of the geometric distance $\| (x,x) \|_2 = \sqrt{2}\, |x|$
from the origin (that is, distance is measured using the $L_2$-norm).  
Thus, rescaling distance in this case by $\sqrt{2}\,n^{-3/4}$
should again collapse the rescaled pressure $n^{19/32}
P_n(x,x)$ to a single curve.  This is seen in figure \ref{figureD}
for values of $n$ ranging from $10$ to $250$.

Comparison to figure \ref{figureB} shows that the data scales
similarly, and may in fact be plotted on the same set of axes.
This is done in figure \ref{figureE} -- that is, the data of 
figures \ref{figureB} and \ref{figureD} are plotted on the
same axes and scale.  This shows that the scaling along the
diagonal is the same as along the $X$-axis.

The data in figure \ref{figureE} is numerical confirmation 
that one may expect the same scaling of the pressure in
any direction from the origin.  This may be made even more
clear by plotting all the data for points $(x,y)$ with
$1\leq x \leq 29$ and $1 \leq y \leq 29$.  That is, plot
the rescaled pressures $n^{19/32} P_n(x,y)$
against $n^{-3/4} \|(x,y)\|_2$ for $4\leq n \leq 250$,
$1\leq x \leq 29$ and $1\leq y \leq 29$
on the same graph.  The result is shown in figure \ref{figureF}.
The data collapse to the same curve.

\begin{figure}[b!]
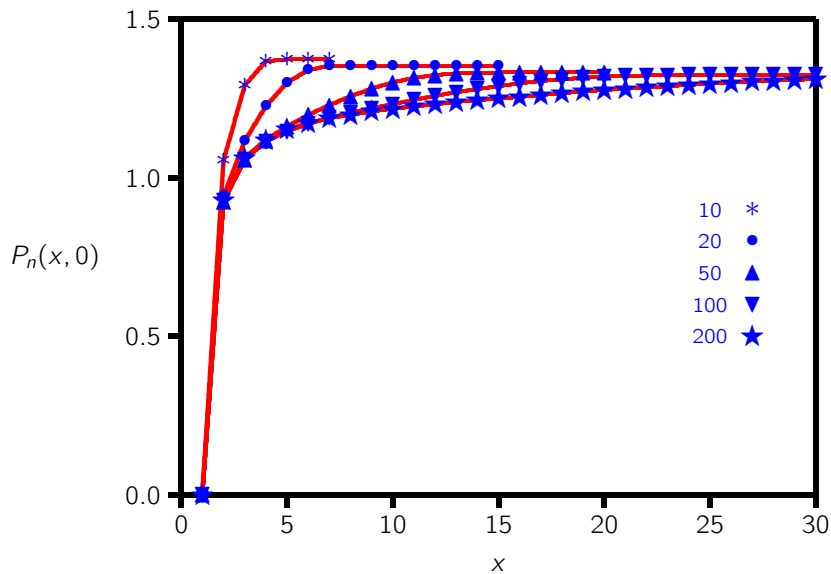

\input figureV.tex
\caption{Velocity of a frictionless unit mass particle accelerating
along the $X$-axis if released at $(1,0)$ for $x=1,2,3,\ldots,30$,
and for $n=10$(\textasteriskcentered), $n=20$($\bullet$),
$n=50$($\blacktriangle$), $n=100$($\blacktriangledown$) and
$n=200$($\bigstar$).
}
\label{figureV}    %%ZXZ[figureV]
\end{figure}

\subsection{The terminal velocity}

The terminal velocity of a unit mass test particle accelerating without
dissipation of energy at constant temperature when released at $(1,0)$
is given by equation \Ref{eqn35}.  Intermediate velocities
as a function of position is given by equation \Ref{eqn34}.

Velocity along the $X$-axis as a function of position is 
plotted in figure \ref{figureV}.  Increasing the size of the polygon
tends to decrease the rate of acceleration and lower intermediate
velocities.  However, the terminal velocity seems to be quite 
insensitive to the length of the polygon, and in each case
levels off close to $1.33$.

\begin{figure}[t!]
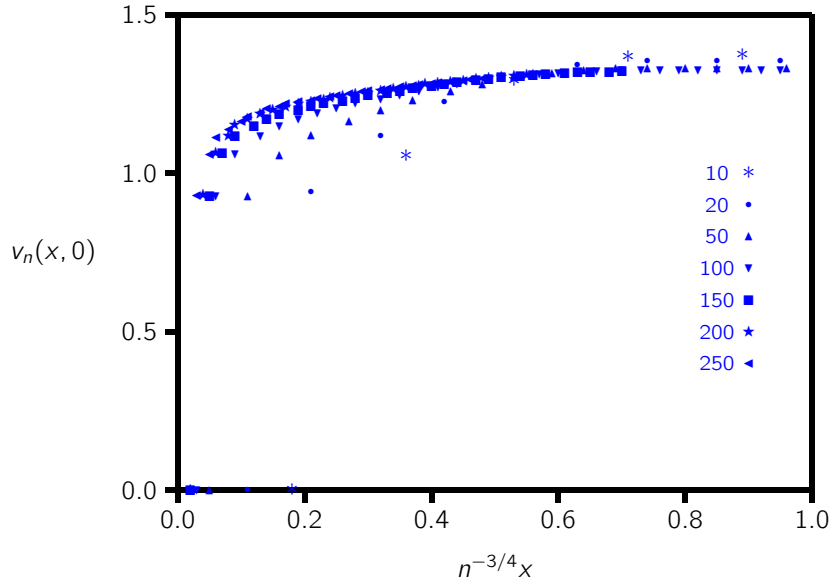

\input figureV2.tex
\caption{Velocity of a frictionless unit mas particle
accelerating along the $X$-axis if released at
$(1,0)$ plotted as a function of rescaled length
$n^{-3/4} x$ for $x=1,2,3,\ldots,30$,
and for $n=10$(\textasteriskcentered), $n=20$($\bullet$),
$n=50$($\blacktriangle$), $n=100$($\blacktriangledown$) and
$n=200$($\bigstar$).
}
\label{figureV2}    %%ZXZ[figureV2]
\end{figure}

\begin{figure}[t!]
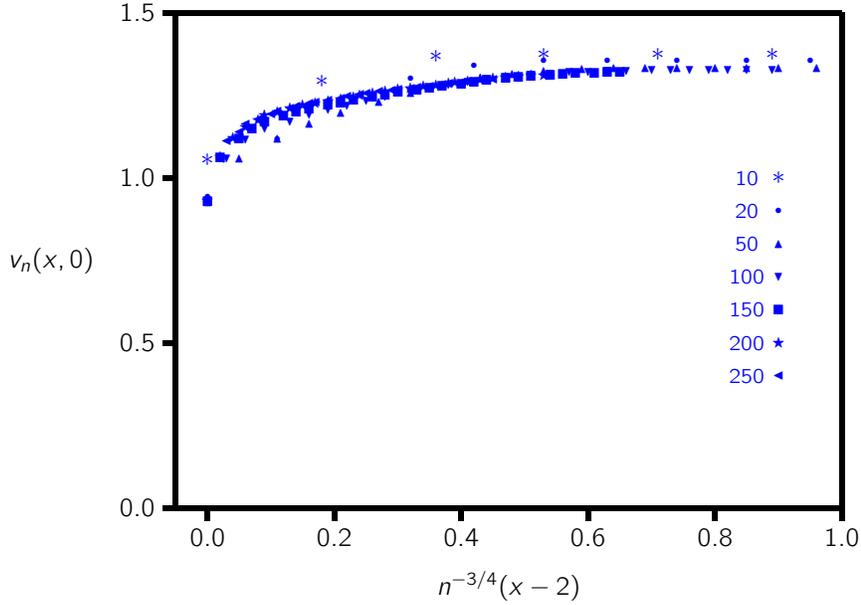

\input figureV3.tex
\caption{Velocity of a frictionless unit mas particle
accelerating along the $X$-axis if released at
$(1,0)$ plotted as a function of rescaled length
$n^{-3/4} (x-2)$ for $x=2,3,4,\ldots,30$,
and for $n=10$(\textasteriskcentered), $n=20$($\bullet$),
$n=50$($\blacktriangle$), $n=100$($\blacktriangledown$) and
$n=200$($\bigstar$).
}
\label{figureV3}    %%ZXZ[figureV3]
\end{figure}

The velocity can be rescaled by rescaling length by $n^\nu$.  In 
figure \ref{figureV2} this is done for all the data in figure
\ref{figureV} and also for data from $n=250$.  With increasing
$n$ the data start to accumulate on a rescaled velocity curve.
This rescaling is compounded by the release of the particle at
$(1,0)$, since the starting point is also rescaled.  By instead
choosing the origin at $x=2$, and then rescaling, the curve
in figure \ref{figureV2} becomes better defined in figure
\ref{figureV3}, with data points clustering more tightly over the
wide range of $n$-values used.

\section{Conclusions}

In this paper the entropic pressure in the vicinity of a polymeric 
molecule was modelled.  There are several other recent papers
which modelled the entropic pressure due to the conformational
entropy of polymer.  These include the pressure of a grafted
two dimensional 
self-avoiding walk on a line \cite{DJ13}, and the entropic pressure
of a two dimensional adsorbing directed path onto the adsorbing line \cite{JvRP13}.

We have used scaling arguments to determine the scaling
of the entropic pressure in the vicinity of a lattice polygon model
of a ring polymer which is rooted at the origin in the square lattice.
The scaling is best described by equation \Ref{Pscaling}.  We tested
this scaling numerically by sampling polygons with the GAS
algorithm, and our data in figures \ref{figureB}, \ref{figureC},
\ref{figureD}, \ref{figureE} and in particular figure \ref{figureF}
provide strong numerical support for equation \Ref{Pscaling}.

In addition, we briefly considered the velocity profile of a unit
mass test particle which accelerates from the polygon do to
the pressure gradient.  Our data suggest that the particle
accelerates to a terminal velocity which is only dependent on the
initial position of the particle, and the intermediate velocities
of the accelerating particle cluster towards a limiting curve
with increasing $n$ if length is rescaled in the model.

The results in this study suggest numerous other questions
regarding the entropic pressure close to polymeric molecules.  The
scaling of a rooted linear polymer can be determined similarly to
the case here, but the situation seems more complicated when
the polymer is grafted onto a hard wall. In three dimensions the
situation seems to be more complicated.  By substituting the
estimates for three dimensional values for the exponents in equation
\Ref{eqnPscaleG}, one would expect that
\begin{equation}
\mathbf{P}_n(a) \simeq \frac{C\, g(a)}{a^6}\, n^{-\tau}
\end{equation}
where $\tau \approx 1.37$, 
for some constant $C$, and where $g(a)$ is given by
equation \Ref{eqn17BC}.  Since this relation is derived using
approximate scaling assumptions, it should be tested numerically
to verify its validity.  In addition, the effects
of knotting on this relation seems a non-trivial and interesting
question.

Finally, observe that if the mean field values
($\gammas1=\sfrac{1}{2}$, $\nu=\sfrac{1}{2}$ and $\alpha_s=0$) of the 
exponents are substituted in equation \Ref{eqnPscaleG}
for $d=4$, then $\mathbf{P}_n(a) \simeq 
\frac{C\, g(a)}{a^8}\, n^{-2}$. 

\vspace{1cm}
\noindent{\bf Acknowledgements:}  The authors are in debt to S.G.
Whittington for comments on the manuscript.  EJJvR acknowledges financial support 
from NSERC (Canada) in the form of a Discovery Grant.

\bibliographystyle{plain}
\bibliography{Farid}

\begin{thebibliography}{10}

\bibitem{AA83}
C.~Aragao~de Carvalho and S.~Caracciolo.
\newblock A new monte-carlo approach to the critical properties of
  self-avoiding random walks.
\newblock {\em Journal de Physique}, 44:323--331, 1983.

\bibitem{BF81}
B.~Berg and D.~Foerster.
\newblock Random paths and random surfaces on a digital computer.
\newblock {\em Physics Letters B}, 106:323--326, 1981.

\bibitem{BMJ00}
T.~Bickel, C.~Marques, and C.~Jeppesen.
\newblock Pressure patches for membranes: The induced pinch of a grafted
  polymer.
\newblock {\em Physical Review E}, 62(1):1124--1127, 2000.

\bibitem{BDD95}
H.D. Bijsterbosch, V.O. De~Haan, A.W. De~Graaf, M.~Mellema, F.A.M. Leermakers,
  M.A. Cohen~Stuart, and A.A. van Well.
\newblock Tethered adsorbing chains: neutron reflectivity and surface pressure
  of spread diblock copolymer monolayers.
\newblock {\em Langmuir}, 11(11):4467--4473, 1995.

\bibitem{BORW05}
R.~Brak, A.L. Owczarek, A.~Rechnitzer, and S.G. Whittington.
\newblock A directed walk model of a long chain polymer in a slit with
  attractive walls.
\newblock {\em Journal of Physics A: Mathematical and General}, 38:4309--4325,
  2005.

\bibitem{C87}
J.L. Cardy.
\newblock Conformal invariance.
\newblock In C.~Domb and J.L. Lebowitz, editors, {\em Phase transitions and
  critical phenomena}, volume~11, pages 55--126, 1983.

\bibitem{CS95}
M.A. Carignano and I.~Szleifer.
\newblock On the structure and pressure of tethered polymer layers in good
  solvent.
\newblock {\em Macromolecules}, 28(9):3197--3204, 1995.

\bibitem{deG79}
P.-G. de~Gennes.
\newblock {\em Scaling concepts in polymer physics}.
\newblock Cornell university press, 1979.

\bibitem{D86}
B.~Duplantier.
\newblock Polymer network of fixed topology: renormalization, exact critical
  exponent $\gamma$ in two dimensions, and d= 4-$\varepsilon$.
\newblock {\em Physical Review Letters}, 57(8):941--944, 1986.

\bibitem{D90}
B.~Duplantier.
\newblock {\em Renormalization and conformal invariance for polymers}.
\newblock SACLAY-SPHT-T-89-162. 1989.

\bibitem{GWF06}
A.~Gholami, J.~Wilhelm, and E.~Frey.
\newblock Entropic forces generated by grafted semiflexible polymers.
\newblock {\em Physical Review E}, 74(4):041803, 2006.

\bibitem{H61}
J.M. Hammersley.
\newblock The number of polygons on a lattice.
\newblock In {\em Proceedings of the Cambridge Philosophical Society},
  volume~57, pages 516--523. Cambridge Univiversity Press, 1961.

\bibitem{HM54}
J.M. Hammersley and K.W. Morton.
\newblock Poor man's monte carlo.
\newblock {\em Journal of the Royal Statistical Society. Series B
  (Methodological)}, pages 23--38, 1954.

\bibitem{HW62}
J.M. Hammersley and D.J.A. Welsh.
\newblock Further results on the rate of convergence to the connective constant
  of the hypercubical lattice.
\newblock {\em The Quarterly Journal of Mathematics}, 13(1):108--110, 1962.

\bibitem{JvR10}
E.J. Janse~van Rensburg.
\newblock Approximate enumeration of self-avoiding walks.
\newblock In {\em Algorithmic Probability and Combinatorics}, volume 520 of
  {\em Contemporary Mathematics}. AMS, 2010.

\bibitem{JvRP13}
E.J. Janse~van Rensburg and T.~Prellberg.
\newblock The pressure exerted by adsorbing directed lattice paths and
  staircase polygons.
\newblock {\em Journal of Physics A: Mathematical and Theoretical},
  46(11):115202, 2013.

\bibitem{JvRR09}
E.J. Janse~van Rensburg and A.~Rechnitzer.
\newblock Generalized atmospheric sampling of self-avoiding walks.
\newblock {\em Journal of Physics A: Mathematical and Theoretical},
  42(33):335001, 2009.

\bibitem{JvRR11}
E.J. Janse~van Rensburg and A.~Rechnitzer.
\newblock Bfacf-style algorithms for polygons in the body-centered and
  face-centered cubic lattices.
\newblock {\em Journal of Physics A: Mathematical and Theoretical},
  44(16):165001, 2011.

\bibitem{JvRR11B}
E.J. Janse~van Rensburg and A.~Rechnitzer.
\newblock Minimal knotted polygons in cubic lattices.
\newblock {\em Journal of Statistical Mechanics: Theory and Experiment},
  2011(09):P09008, 2011.

\bibitem{JvRR11A}
E.J. Janse~van Rensburg and A.~Rechnitzer.
\newblock On the universality of knot probability ratios.
\newblock {\em Journal of Physics A: Mathematical and Theoretical},
  44(16):162002, 2011.

\bibitem{DJ13}
I.~Jensen, W.G. Dantas, C.M. Marques, and J.F. Stilck.
\newblock Pressure exerted by a grafted polymer on the limiting line of a
  semi-infinite square lattice.
\newblock {\em Journal of Physics A: Mathematical and Theoretical},
  46(11):115004, 2013.

\bibitem{MS93}
N.~Madras and G.~Slade.
\newblock {\em The self-avoiding walk}.
\newblock Birkha\"user, 1993.

\bibitem{MN91}
C.M. Mate and V.J. Novotny.
\newblock Molecular conformation and disjoining pressure of polymeric liquid
  films.
\newblock {\em The Journal of Chemical Physics}, 94:8420--8437, 1991.

\bibitem{MTW77}
K.M. Middlemiss, G.M. Torrie, and S.G. Whittington.
\newblock Excluded volume effects in the stabilization of colloids by polymers.
\newblock {\em The Journal of Chemical Physics}, 66:3227--3232, 1977.

\bibitem{N82}
B.~Nienhuis.
\newblock Exact critical point and critical exponents of $o(n)$ models in two
  dimensions.
\newblock {\em Physical Review Letters}, 49:1062--1065, 1982.

\bibitem{N84}
B.~Nienhuis.
\newblock Coulomb gas description of 2-d critical behaviour.
\newblock {\em Journal of Statistical Physics}, 34:731--761, 1984.

\bibitem{P91}
P.~Pincus.
\newblock Colloid stabilization with grafted polyelectrolytes.
\newblock {\em Macromolecules}, 24(10):2912--2919, 1991.

\end{thebibliography}

\end{document}